\documentclass[1p]{elsarticle}

\usepackage{mathrsfs}
\usepackage{amssymb,amsthm}
\usepackage{amsmath,amsfonts}
\usepackage{mathtools}
\usepackage{caption}
\usepackage{xcolor}
\usepackage{comment}
\usepackage{lipsum}
\usepackage{graphicx}
\usepackage{epstopdf}
\usepackage{algorithmic}
\usepackage{bbold}
\usepackage{tikz}
\usepackage{subfig}
\usepackage{adjustbox}
\usepackage[noresetcount,vlined,ruled]{algorithm2e} 
\usepackage[compact]{titlesec}
\usepackage{autobreak} 
\usepackage{hyperref}
\allowdisplaybreaks

\newcommand{\myket}[1]{\lvert #1 \rangle}
\usepackage{tikz}
\usetikzlibrary{arrows}
\newcommand{\psib}{\boldsymbol{\psi}}
\newcommand{\alphab}{\boldsymbol{\alpha}}

\ifpdf
  \DeclareGraphicsExtensions{.eps,.pdf,.png,.jpg}
\else
  \DeclareGraphicsExtensions{.eps}
\fi

\theoremstyle{remark}

\journal{SIAM Journal of Control and Optimization}

\begin{document}
\begin{frontmatter}

\title{Binary Optimal Control Of Single-Flux-Quantum Pulse Sequences}

\author[llnl]{Ryan H. Vogt}
\ead[email]{vogt9@llnl.gov}

\author[llnl]{N. Anders Petersson}
\ead[email]{petersson1@llnl.gov}

\cortext[cor1]{Corresponding author}
\address[llnl]{Center for Applied Scientific Computing, Lawrence Livermore National Laboratory, Livermore, CA 94550, USA.}

\renewcommand{\thefootnote}{\arabic{footnote}}

\begin{abstract}
We introduce a binary, relaxed gradient, trust-region method for optimizing pulse sequences for single flux quanta (SFQ) control of a quantum computer.
The pulse sequences are optimized with the goal of realizing unitary gate transformations. Each pulse has a fixed amplitude and duration. We model this process as an binary optimal control problem, constrained by Schr\"{o}dinger's equation, where the binary variables indicate whether each pulse is on or off. We introduce a first-order trust-region method, which takes advantage of a relaxed gradient to determine an optimal pulse sequence that minimizes the gate infidelity, while also suppressing leakage to higher energy levels. The proposed algorithm has a computational complexity of ${\cal O}(p\log(p))$, where $p$ is the number of pulses in the sequence. We present numerical results for the H and X gates, where the optimized pulse sequences give gate fidelity's better than $99.9\%$, in $\approx 25$ trust-region iterations. 
\end{abstract}


\end{frontmatter}

\section{Introduction} \label{s:intro}

Over the past decades the development of algorithms for a quantum computer, which differ from algorithms for classical computers, and need particular special treatment, brought a great practical interest to quantum computers \cite{cleve1998quantum, montanaro2016quantum}. Several high impact problems such as combinatorial optimization \cite{han2000genetic}, quantum cryptography\cite{mavroeidis2018impact}, and large dimensional linear systems \cite{nakahara2008quantum} are a few notable examples that motivate the interest in developing quantum computing technologies. 

Quantum algorithms become attractive when a quantum computer has a large amount of qubits, however, state of the art quantum computers currently only have on the order of hundreds of qubits; and these qubits do not support error correction. Many quantum algorithms demand at least one million to one-hundred million qubits \cite{mcdermott2018quantum} to be an attractive alternative to classical computing. A popular approach to building a quantum computer is taking advantage of superconducting electronic circuits, which is referred to as superconducting quantum computing. While the superconducting circuits reside inside a dilution refrigerator at milli-Kelvin temperatures, conventional qubit control pulses consisting of modulations of a microwave carrier tone, are generated outside the fridge at room temperature. As a result, the control waveform that is delivered to the qubit is the convolution of the applied waveform with the transfer function of the wiring into the refrigerator. This transfer function is known to depend on the frequency of the carrier wave, but is hard to characterize precisely. An additional challenge from having separate control lines for each qubit is the significant heat load on the milli-Kelvin environment inside the dilution fridge, see~\cite{li2019hardware} for further discussions.

An alternative approach to controlling an array of qubits is by using a classical co-processor that resides inside the fridge. This approach to coherent control involves irradiation of the qubit with trains of quantized flux pulses derived from the single-flux-quantum (SFQ) digital logic family \cite{Likharev-91}. Here, classical bits of information are stored as the presence or absence of a phase slip across a Josephson junction in a given clock cycle, where each phase slip corresponds to a voltage pulse whose time integral equals $\Phi_0 = h/2e$, i.e., the superconducting flux quantum.
These circuits have gained the attention of the quantum computing community for several reasons: (i) SFQ circuits are a promising candidate for scalable quantum computing, i.e many qubit systems; and (ii) SFQ circuits can realize gate and measurement fidelity to high accuracy \cite{mcdermott2018quantum}. SFQ circuits offer one possible solution to realizing these many qubit systems. One limitation of the SFQ technology is that calculations are done by applying a sequence of pulses of fixed amplitude. This means to conduct computations on a SFQ quantum computer pulses must be administered in a particular manner to achieve desired unitary gate transformations.

The study of controlling quantum mechanical systems in an optimal manner, referred to as quantum optimal control theory, has increasingly become an important tool for developing optimal control strategies that yield a favorable outcome in a quantum mechanical system. The performance of a control strategy can be gauged by a preferred metric, such as realizing gates to high fidelity. We refer the reader to \cite{glaser2015training} and the references therein, for a review of the current state-of-the-art of quantum optimal control in different domains, such as  atomic, molecular, and chemical physics, magnetic resonance, and quantum information and communication. We make special note that \cite{niu2019universal}, and the references therein, discusses the state of the art approach for solving quantum optimal control problems using reinforcement machine learning. Reinforcement machine learning is attractive in the quantum computing setting because unlike supervised machine learning, there is no explicit need to have training data. Generating data, even for established quantum technologies, could still require expensive experiments to be conducted. Some quantum architectures that are of interest to researches are still theoretical so data cannot be gathered in those instances. We include this reference so that i) the reader is aware of the effectiveness machine learning has had on the field of quantum optimal control for realizing robust high fidelity one and two qubit gates; and ii) while machine learning approaches for solving the SFQ problem is out of the scope of this paper, we wish to not dismiss the idea that machine learning approaches could yield effective control strategies for the SFQ problem. We favor leveraging the binary structure that appears in the problem to develop an specialized trust-region method for the SFQ architecture that yields quality control strategies quickly.

In this paper we introduce a binary optimal control problem to determine the manner in which to apply pulse sequences to the SFQ circuit so that gates are realized to high fidelity. In this work we focus on controlling only a single qubit. We model the control as a binary vector of fixed size, where each element of the binary vector represents the decision to apply a pulse, or no to apply a pulse, to the SFQ circuit. Note that the SFQ architecture only provides two control options in each time interval: either apply a pulse with predefined amplitude and duration, or not apply a pulse, i.e., a pulse with zero amplitude. Since we are deciding on a collection of fixed amplitude pulses to realize a gate, rather than a pulse in which the amplitude varies in time, the popular GRAPE \cite{khaneja2005optimal} algorithm is not directly applicable. 
To solve the optimal control problems with binary variables there are several standard approaches: genetic algorithms \cite{ga1,ga2,ga3}, the branch and bound algorithm \cite{b2,b1}, and annealing \cite{nahar1986simulated}. Genetic algorithms have already been utilized for finding pulse sequences in the SFQ problem~\cite{mcdermott2018quantum}, but a quality solution is not guaranteed. To the best of our knowledge annealing has not been applied to the SFQ problem, but like genetic algorithms, would not guarantee a quality solution. Branch and bound on the other hand does offer a notion of finding a quality solution of the binary optimal control problem. However in practice a large amount of binary variables will make the method computationally intractable. In this paper we introduce a first-order trust-region method that is computationally tractable, as exemplified by optimizing pulse sequences for realizing H and X gates. 

The paper is outlined as follows: In Section \ref{s:2} we introduce a truncated modal expansion for Schr\"odinger equation to model a quantum system of one qubit, and the Hamiltotian we use to model a SFQ quantum circuit using SFQ pulses. In Section \ref{s:adjointgrad} we introduce the optimal control problem for the SFQ quantum circuit; here, we define the objective function of the optimal control problem that is the sum of an gate infidelity term and a leakage suppression term. This objective function balances the interests of realizing gates to high infidelity, while also suppressing leakage to the guard states. In Section \ref{s:trustregion} we introduce a first-order trust-region method to solve the SFQ optimal control problem. In Section \ref{s:experiements} we solve the optimal control problem to realize two quantum gates to high fidelity: the H and X gate. In addition, we explore how the tip angle, which we define in the paper, impacts gate realization. In Section \ref{s:conclusions} we summarize our findings.
 

\section{SFQ optimal control problem}\label{s:2} \label{s:problem}
\begin{figure}[htbp]
\centering
\includegraphics[width=0.45\textwidth,height=0.25\textheight]{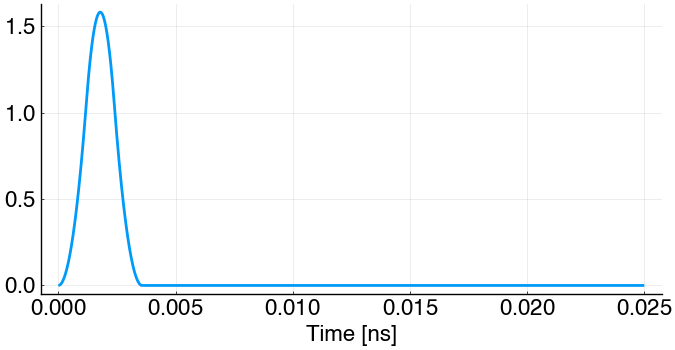}
\caption{Illustration of a B-spline pulse function with SFQ time step $\tau_p=2.5\times 10^{-2}$ ns and pulse duration $\delta = 4\times 10^{-3}$ ns. }
\label{fig:spline}
\end{figure}

Consider a truncated modal expansion of the Schr\"odinger equation with $N>0$ states for realizing a gate transformation. The truncation of the expansion is justified by discouraging population of the states with the highest energy levels. The gate transformation is defined in the "essential" subspace corresponding to the $E>0$ lowest energy levels and we let $G = N - E \geq 0$ denote the number of ``guard'' states. The evolution of the state vector satisfies $\psib(t;\alphab) = U(t;\alphab)\psib(0;\alphab)$, where the $N\times N$ complex-valued solution operator matrix $U(t;\alphab)$ satisfies Schr\"odinger's equation in matrix form:
\begin{equation}\label{eq:schrodinger_matrix}
\frac{d U}{d t} + \frac{i}{\hbar} H(t;\boldsymbol{\alpha}) U = 0, \quad t \in \Omega=[0,T], \quad U(0) = I_{N}.
\end{equation}
Here, $\hbar$ is the reduced Plank constant, $i=\sqrt{-1}$ is the imaginary unit, and $T>0$ is the duration of the gate transformation.
The $N\times N$ identity matrix is denoted $I_{N}$ and represents the canonical basis for the $N$-dimensional  state vector. The  Hamiltonian matrix is $H(t;\boldsymbol{\alpha})$, in which the time-dependence is parameterized by the $p$-dimensional binary control pulse vector: $\boldsymbol{\alpha} \in \{0,1\}^p, \; p\geq 1$. As a result, the solution operator matrix $U$ depends implicitly on $\alphab$.
The Hamiltonian matrix is assumed to be of the form
\begin{align*}
    H(t) = H_0 + H_{SFQ}(t),
\end{align*}
where the system  Hamiltonian is modeled by
\begin{align*}
\frac{H_0}{\hbar} &= \omega a^\dagger a - \frac{\xi}{2} a^\dagger a^\dagger a a.
\end{align*}
Here, $a$ and $a^\dagger$ are the lowering and raising operators. The constants $\omega_a$ and $\xi$ represent the qubit's fundamental frequency and anharmonicity (self-Kerr coefficient) respectively. The control Hamiltonian is given by
\begin{align*}
    H_{SFQ}(t) = \frac{C_c}{C} v(t) \hat{Q},\quad \hat{Q} = 2e \, i(a-a^\dagger)
\end{align*}
Here, $C_c$ and $C$ are the coupling and Josephson capacitance, respectively, and $e$ is the electron charge constant. We divide time into pulse intervals of length $\tau_p>0$, which we refer to as the SFQ time steps, with $t_k= k \tau_p$, for $k=0,1,\ldots,p$, such that $T=p\tau_p$. We model $v(t)$, which represents the voltage of the pulse delivered to the system in the following manner:
\begin{align*}
    v(t) = \sum_{k=1}^p \alpha_k \hat{B}(t - t_{k-1}), \quad t\in[t_{k-1}, t_{k}].
\end{align*}
We use a quadratic B-Spline basis function \cite{gordon1974b,anders2020discrete} $\hat{B}(t)$ to represent the pulse delivered at each SFQ time step. Here, $[\boldsymbol{\alpha}]_k=\alpha_{k}$ for $k=1,..,p$ is a binary variable that indicates whether the pulse is on or off during the interval $t\in[t_{k-1}, t_k]$,
\[
\alpha_k = \begin{cases}
1,\quad & \mbox{pulse on}\\
0,\quad & \mbox{pulse off}.
\end{cases}
\]
If the pulse is on, the function $v(t)$ is scaled such that its strength corresponds to one magnetic flux quanta,
\begin{align*}
    \int_{t_k}^{t_k+\tau_p} v(t) \, dt = \Phi_0.
\end{align*}
It is convenient to scale the time-function such that
\begin{align*}
    \Tilde{v}(t) = \Phi_0 v(t).
\end{align*}
Because the magnetic flux quanta is defined by $\Phi_0 = 2\pi \hbar/(2 e)$,
\begin{align*}
   \frac{C_c}{C} \Phi_0 v(t) 2e\, i (a-a^\dagger) =
   \hbar \beta \Tilde{v}(t)\, i (a - a^\dagger),\quad \beta = 2\pi \frac{C_c}{C}.
\end{align*}
Thus,
\begin{align*}
    \frac{1}{\hbar}H_{SFQ}(t) = \beta \Tilde{v}(t) \, i(a-a^\dagger).
    %
\end{align*}

\begin{align*}
\Tilde{v}(t) = \frac{1}{\gamma}\sum_{k=1}^p \alpha_k \hat{B}(t - t_{k-1}), \quad t\in[t_{k-1}, t_{k}],\quad \gamma = \int_{0}^{\delta} \hat{B}(t)\,dt.
\end{align*}
Here  $0 < \delta \leq \tau_p$ is the pulse duration, which defines the support of $\hat{B}(t)$. By defining $\tilde{v}(t)$ in this manner we ensure that
\begin{align*}
    \int_{t_{k-1}}^{t_k} \Tilde{v}(t)dt=1,
\end{align*} if the pulse is on, which corresponds to applying a pulse with strength $\beta$. We illustrate a single pulse in Figure \ref{fig:spline} with SFQ time step $\tau_p=2.5\times 10^{-2}$ nanoseconds (ns)  and pulse duration $\delta = 4\times 10^{-3}$ ns. 

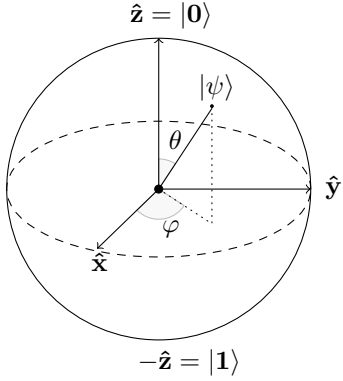
\begin{figure}[htbp]
\centering
\begin{tikzpicture}[line cap=round, line join=round]
  \clip(-2.19,-2.49) rectangle (2.66,2.58);
  \draw [shift={(0,0)}, lightgray, fill, fill opacity=0.1] (0,0) -- (56.7:0.4) arc (56.7:90.:0.4) -- cycle;
  \draw [shift={(0,0)}, lightgray, fill, fill opacity=0.1] (0,0) -- (-135.7:0.4) arc (-135.7:-33.2:0.4) -- cycle;
  \draw(0,0) circle (2cm);
  \draw [rotate around={0.:(0.,0.)},dash pattern=on 3pt off 3pt] (0,0) ellipse (2cm and 0.9cm);
  \draw (0,0)-- (0.70,1.07);
  \draw [->] (0,0) -- (0,2);
  \draw [->] (0,0) -- (-0.81,-0.79);
  \draw [->] (0,0) -- (2,0);
  \draw [dotted] (0.7,1)-- (0.7,-0.46);
  \draw [dotted] (0,0)-- (0.7,-0.46);
  \draw (-0.08,-0.3) node[anchor=north west] {$\varphi$};
  \draw (0.01,0.9) node[anchor=north west] {$\theta$};
  \draw (-1.01,-0.72) node[anchor=north west] {$\mathbf {\hat{x}}$};
  \draw (2.07,0.3) node[anchor=north west] {$\mathbf {\hat{y}}$};
  \draw (-0.5,2.6) node[anchor=north west] {$\mathbf {\hat{z}=|0\rangle}$};
  \draw (-0.4,-2) node[anchor=north west] {$-\mathbf {\hat{z}=|1\rangle}$};
  \draw (0.4,1.65) node[anchor=north west] {$|\psi\rangle$};
  \scriptsize
  \draw [fill] (0,0) circle (1.5pt);
  \draw [fill] (0.7,1.1) circle (0.5pt);
\end{tikzpicture}
\caption{Illustration of the Bloch sphere. }
\label{fig:bloch}
\end{figure}

For the case of $E=2$ essential states in a system, the Bloch sphere provides a convenient geometric representation of singe qubit transformations~\cite{sakurai1995modern}. We illustrate the Bloch sphere in Figure \ref{fig:bloch}. Starting from the ground state (represented by the North pole on the Bloch sphere) one SFQ pulse will perturb the state away from the North pole. In terms of spherical coordinates, the corresponding change in polar angle will be referred to as the tip angle $\theta$, which is related to the pulse strength by $\beta = \theta/\pi \tau_p$. 

\section{Objective and gradient computation}\label{s:adjointgrad}
In this section we introduce the objective function for our optimal control problem. We use this objective function to write a closed form expression for the gradient with respect to the binary control vector $\alphab$. 

In this work we denote the column vectors of the solution operator matrix by 
\[
U = \begin{bmatrix} \psib_0 & \psib_1 & \cdots & \psib_{N-1}\end{bmatrix}\in\mathbb{C}^{N\times N}, \quad%
\psib_j = \begin{bmatrix}
\psi_{0,j}\\
\psi_{1,j}\\
\vdots\\
\psi_{N-1,j}
\end{bmatrix}\in\mathbb{C}^{N\times 1}.
\]
The unitary target gate transformation $V_E$ is only defined for the first $E$ rows and columns of $U(T,\boldsymbol{\alpha})$. For notional simplicity we embed it in the blocked matrix $V$:
\begin{equation*}
V = \begin{bmatrix}
V_E & \boldsymbol{0}_{E\times G} \\
\boldsymbol{0}_{G\times E} & \boldsymbol{0}_{G\times G}
\end{bmatrix}\in\mathbb{C}^{N\times N}.
\end{equation*}
The overlap between the solution operator matrix and the target gate matrix can be measured by the gate infidelity:
\begin{equation}\label{eq:objf}
  {\cal J}_1(U_T(\boldsymbol{\alpha})) = 1 - \frac{1}{E^2} \left| \left\langle U_T(\boldsymbol{\alpha}), V \right\rangle_F  \right|^2 \geq 0, \quad
    U_T(\boldsymbol{\alpha}) := U(T,\boldsymbol{\alpha}),
\end{equation}
where $\langle \ \cdot \ , \ \cdot \ \rangle_F$ denotes the Fr\"{o}benius matrix scalar product (with associated norm $\|\cdot\|_F$).  Because the $G$ last rows and columns of $V$ are zero, only the first $E$ rows and columns of $U_T$ matter for this part of the objective function. We refer to $\mathcal{J}_1$ as the infidelity term in the remainder of this paper. In this work we strive to construct pulse sequences that cause the infidelity to be below $10^{-3}$. This is equivalent to requiring the gate fidelity to be larger than $99.9\%$.

An important observation we make is that $\alpha_k$ is either zero or one for $k=1,\ldots,p$. This means that we only need to solve Schr\"odinger's equation \eqref{eq:schrodinger_matrix} for the duration of one SFQ time step in two cases: when the pulse is on, or when it is off. We denote by $D_0$ the solution corresponding to when the pulse is off, and by $D_1$ when the pulse is on. We can then represent the solution of the state equation at the final time, $t=T$, as
\begin{align*}\label{sfqpulse}
    U(T,\boldsymbol{\alpha}) = D_{\alpha_p}D_{\alpha_{p-1}}...D_{\alpha_1},
\end{align*}
which also implies
\begin{equation}\label{eq_disc-sol}
    U_j = U(t_j)= 
    \begin{cases}
    D_{\alpha_j} D_{\alpha_{j-1}}\cdots D_{\alpha_1},& j\geq 1,\\
    I, & j=0.
    \end{cases}
\end{equation}
This is a helpful observation because it means that, based on the unitary matrices $D_0$ and $D_1$, the solution operator matrix can be evaluated by $p$ matrix multiplications, where $p$ is the number of SFQ pulses. Furthermore, the matrix $D_0$ can be calculated by matrix exponentiation. Hence, only the matrix $D_1$ needs to be calculated through numerical time stepping. This only needs to be performed once, in a pre-processing stage of the algorithm.

The population of the guard states can be measured by evaluating the following expression:
 \begin{equation}\label{eq:objf-guard}
   \frac{1}{T} \int_0^T \sum_{j=0}^{E-1}
    \left\langle \boldsymbol{\psi}_j(t,\boldsymbol{\alpha}), W \boldsymbol{\psi}_j(t,\boldsymbol{\alpha}) \right\rangle_2 \, dt =
    \frac{1}{T} \int_0^T \left\langle UP, WUP\right\rangle_F,
 \end{equation}
 where the rectangular matrix
\begin{align*}
    P = \begin{bmatrix}
    I_E \\
    0_G
    \end{bmatrix}\in\mathbb{R}^{N\times E}
\end{align*}
is used to single out the first $E$ columns of the matrix that is applied to it. Further, $W\in \mathbb{R}^{N\times N}\geq 0$ is a diagonal positive semi-definite weight matrix where only the rows of $W$ corresponding to the guard states have positive elements.  For example, if $E=2$ and $G=2$, the weight matrix may be
\begin{equation*}\label{eq_wmat}
W = \begin{bmatrix}
0 &0 &0 &0 \\
0& 0 &0 &0 \\
0& 0 & w_1  & 0\\
0& 0 & 0  & w_2
\end{bmatrix},\quad w_1,w_2 \geq 0.
\end{equation*}
We approximate \eqref{eq:objf-guard} with the trapezoidal rule, using $\tau_p$ as the time step, which results in the following expression:
\begin{align}\label{eq_g2}
    {\cal J}_2 = \frac{1}{p} \left(
    \frac{1}{2}\langle U_0 P,W U_0  P\rangle_F + \sum_{j=1}^{p-1} \langle U_j P, W U_j P\rangle_F + \frac{1}{2}\langle U_p P,W U_p  P\rangle_F
    \right).
\end{align}
Here, $U_j$ is defined by \eqref{eq_disc-sol}. In the following, we refer to $\mathcal{J}_2$ as the leakage term.

For the quantum control problem with guard states, we formulate the optimization problem as
\begin{gather*}
  \underset{\boldsymbol{\alpha}}{\text{min}}\,  {\cal J}(\boldsymbol{\alpha}) := {\cal
    J}_1(U_T(\boldsymbol{\alpha})) + C_1{\cal J}_2(U(\cdot,\boldsymbol{\alpha})), \quad C_1> 0, \label{eq_ineq-constraints2}
\end{gather*}
where $U(t,\boldsymbol{\alpha})$ is governed by \eqref{eq:schrodinger_matrix}. From numerical experimentation we find that $C_1=10^{-2}$
helps insure that both terms of $\mathcal{J}$ are of the same order of magnitude during the optimization. We note that this optimization problem is a large dimensional nonlinear integer programming problem (NIP), since $\boldsymbol{\alpha}$ is a binary (integer) vector, ${\cal J}(\boldsymbol{\alpha})$ is nonconvex. NIP's are typically in the class of computations problems that are non-polynomial (NP), which is to say that there does not currently exist a polynomial time algorithm to solve them. This causes large dimension NIP's to be extremely hard to solve in practice. We discuss more about NIP's in Section \ref{s:trustregion}, and propose a trust-region algorithm to solve the NIP that arises from the SFQ optimal control problem. An overview of the state-of-the-art theory, and numerical solutions for NIP's can be read in \cite{nl2,nl3,lee2011mix,nl1}.

Next, we discuss the gradient computation of the objective function. We note that,
\begin{equation*}\label{gradeq}
 \frac{\partial {\cal J}}{\partial\alpha_k} =  
 \frac{\partial {\cal J}_1}{\partial\alpha_k} + 
 C_1\frac{\partial {\cal J}_2}{\partial\alpha_k}.
\end{equation*}
In this work we define $A_{j} = D_{\alpha_j}$ and $B_j = \partial D_{\alpha_j}/\partial \alpha_j$. The latter expression is defined in the sense of a relaxed gradient that is calculated by temporarily assuming that $\alpha_j$ is real, with $0\leq\alpha_j \leq 1$. The continuous gradient can then be defined and evaluated at its upper and lower bounds, corresponding to the binary variable $\alpha_j\in \{0,1\}$. In the following, the gradient of the objective function is to be understood in the sense of this relaxed gradient.

In Appendix \ref{s:gradcomp}, we derive the expressions for the gradients of $\mathcal{J}_1$ and $\mathcal{J}_2$. Here we only present the main results.
To calculate the gradient of the infidelity term $\mathcal{J}_1$, we first note that \eqref{eq:objf} can be written
\begin{align}\label{eq_J1}
    \mathcal{J}_1 = 1 - \frac{1}{E^2} S_T \bar{S}_T,\quad
    S_T = \langle U_p P, V P \rangle_F,
\end{align}
where $\bar{S}_T$ denotes the complex conjugate of $S_T$.
The gradient can then be expressed as
\begin{align*}
    \frac{d {\cal J}_1}{d \alpha_{p-q}} = -\frac{2}{E^2} \mbox{Re} \left( \bar{S}_T \left\langle B_{p-q} U_{p-q-1} P, \Lambda_{p-q} P\right\rangle_F 
    \right), \quad q=0,\ldots,p-1,
\end{align*}
where the discrete adjoint variable $\Lambda_k$ satisfies
\begin{align*}
   \Lambda_{p-q} = \begin{cases}
    \Lambda_p = V ,& q=0,\\
        \Lambda_{p-q}  = A^\dagger_{p-q+1} \Lambda_{p-q+1},&q = 1,\ldots,p-1.
    \end{cases}
\end{align*}

The gradient of the leak term can be expressed as
\begin{align*}\label{eq_adj-grad-e}
    \frac{d{\cal J}_2}{d\alpha_{p-q}} = 
    \frac{2}{p} \mbox{Re} 
     \left\langle 
     B_{p-q}U_{p-q-1}P, \Tilde{\Lambda}_{p-q}P \right\rangle_F,\quad q=0,\ldots, p-1.
\end{align*}
In this case, the discrete adjoint variable is calculated according to
\begin{align*}
    \Tilde{\Lambda}_{p-q} = \begin{cases}
    \frac{1}{2} W U_p P,& q=0,\\
        W U_{p-q}P + A^\dagger_{p-q+1} {\widetilde{\Lambda}}_{p-q+1},&q = 1,2,\ldots,p-1.
    \end{cases}
\end{align*}

\section{Trust-region method}\label{s:trustregion}
Trust-region methods are a popular approach for solving continuous nonlinear constrained optimization problems  \cite{nocedal.wright:99}. Given an initial feasible solution to the constrained optimization problem at iteration $k$, a first or second-order Taylor series approximation is written 
for  an objective function $f(\boldsymbol{x})$ to be minimized i.e 
\begin{align*}
    f(\boldsymbol{x}) \approx \tilde{f}(\boldsymbol{x}) =  f(\boldsymbol{x}_k) + \nabla_{\boldsymbol{x}}^Tf(\boldsymbol{x}_k)(\boldsymbol{x} -\boldsymbol{x}_k),
\end{align*}
or
\begin{align*}
  f(\boldsymbol{x}) \approx \tilde{f}(\boldsymbol{x}) =  f(\boldsymbol{x}_k) + \nabla_{\boldsymbol{x}}^Tf(\boldsymbol{x}_k)(\boldsymbol{x} -\boldsymbol{x}_k) + \frac{1}{2}(\boldsymbol{x} -\boldsymbol{x}_k)^T \nabla_{\boldsymbol{x}}^2f(\boldsymbol{x}_k)(\boldsymbol{x} -\boldsymbol{x}_k)  ,
\end{align*}
where $\nabla_{\boldsymbol{x}}f$ and $\nabla_{\boldsymbol{x}}^2f$ are the gradient and hessian respectively. Next a trust-region subproblem is solved:
\begin{gather*}
  \underset{\boldsymbol{x}}{\text{min}}\, \tilde{f}(\boldsymbol{x}) \\
  \text{subject to }\\
  ||\boldsymbol{x}-\boldsymbol{x}_k||_2 \leq \Delta_k \\
  g(\boldsymbol{x})=0\\
  h(\boldsymbol{x}) \leq 0,
\end{gather*}
where $\Delta_k$ is referred to as the trust-region radius.
Once the solution $\boldsymbol{x}$ is found, a condition is tested to evaluate if the truncated Taylor series approximation made is valid to the true objective function $f$ constrained to the trust-region radius $\Delta_k$. In the case that the assumption is valid, the trust-region radius is increased by some rule in order to attempt to globalize the search for a local minimum since the approximation of the objective function and the objective function are approximately the same locally. If the assumption is not valid, then the trust-region radius is decreased by some rule. This indicates that the approximation is not accurate to the true objective function within the trust-region radius. This process continues until the trust-region radius is small, indicating that a local minimum has been found, thus terminating the trust-region method.

In \cite{vogt,vogt2021solving} a trust-region method was introduced for optimal control problems with binary variables. The main idea of this algorithm is use the standard first-order trust-region method suggested in \cite{nocedal.wright:99} for solving nonlinear optimization problems. In this algorithm the gradient is understood to be the relaxed gradient, derived from the continuous relaxation of the binary variable as explained above. 

For completeness, we present the first-order trust-region method in Algorithm \ref{A:Steepest}.
\begin{algorithm}[htbp]
\SetAlgoVlined
\caption{Steepest-Descent Trust-Region Algorithm.\label{A:Steepest}}
Given an initial trust-region radius $\Delta_{0} \geq 1$, and an initial parameter vector $\boldsymbol{\alpha}^{(0)} \in \{0,1\}^{p}$.

Select an acceptance step parameter $\hat{\rho}$, for example $\hat{\rho}=0.75$, and initialize $k = 0$.

Evaluate the objective function $\mathcal{J}_{}^{(k)} = \mathcal{J}^{}(\boldsymbol{\alpha}^{(k)})$ and the gradient $g_{}^{(k)} = \nabla_{\boldsymbol{\alpha}} \mathcal{J}^{}(\boldsymbol{\alpha}^{(k)})$.

\While{$\Delta_k \geq 1$}{
  Solve the trust-region (knapsack) sub-problem for $\widehat{\boldsymbol{\alpha}}$:
  \begin{align*}
    \widehat{\boldsymbol{\alpha}} = \underset{\boldsymbol{\alpha}}{\text{minimize}} \quad  &g_{}^{(k)^T}\left(  \boldsymbol{\alpha} -\boldsymbol{\alpha}^{(k)} \right) + \mathcal{J}_{}^{(k)} &  \\ 
    \quad & \|\boldsymbol{\alpha}-\boldsymbol{\alpha}^{(k)}\|_1 \leq \Delta_k  &   \\
    &  \boldsymbol{\alpha}\in \{0,1 \}^{p} & 
  \end{align*}
  Solve the state equation for the binary parameter vector $\widehat{\boldsymbol{\alpha}}$ and evaluate the objective $\mathcal{J}_{}(\widehat{\boldsymbol{\alpha}}))$.
  
  Compute the ratio of actual over predicted reduction:   \begin{align*}
    \rho_k = \frac{\mathcal{J}_{}^{(k)} - \mathcal{J}(\widehat{\boldsymbol{\alpha}})}{-\big( g_{}^{(k)} \big)^T \big( \widehat{\boldsymbol{\alpha}} - \boldsymbol{\alpha}^{(k)}\big)}
    \end{align*}

 \uIf{$||g^{(k)}||_2=0$ or $\hat{\alphab} = \alphab^{(k)}$}{ $\alphab^{(k)}$ is a local solution\\
 Set $\alphab^{(k+1)} = \alphab^{(k)} $\\
 Reduce to trust-region radius $\Delta_{k+1}=0$

}
  \uElseIf{$\rho_k > \hat{\rho}$}{
    Accept the step: $\boldsymbol{\alpha}^{(k+1)} = \widehat{\boldsymbol{\alpha}}$, and evaluate the gradient  $g_{}^{(k+1)} = \nabla_{\hat{\alphab}}\mathcal{J}(\boldsymbol{\alpha}^{(k+1)})$\\
    \uIf{$\| \boldsymbol{\alpha}^{(k+1)} - \boldsymbol{\alpha}^{(k)} \|_1 = \Delta_k$,}{
    	Increase the trust-region radius $\Delta_{k+1} = 2\Delta_{k}$}	
  }
  \uElseIf{$\rho_k >0$}{
    Accept the step $\boldsymbol{\alpha}^{(k+1)}= \widehat{\boldsymbol{\alpha}}$, and evaluate the gradient $g_{}^{(k+1)} = \nabla_{\hat{\alphab}}\mathcal{J}(\boldsymbol{\alpha}^{(k+1)})$\\
    Keep trust-region radius unchanged $\Delta_{k+1} = \Delta_k$
  }
  \Else{
    Reject the step, set $\boldsymbol{\alpha}^{(k+1)} = \boldsymbol{\alpha}^{(k)}$, and copy the gradient $g_{}^{(k+1)} = g_{}^{(k)}$\\
    Reduce the trust-region  radius $\Delta_{k+1} = \text{floor}\left(\frac{\Delta_k}{2}\right)$
  } 

  Set $k \gets k+1$ 
} 
\end{algorithm}
We note that because $\boldsymbol{\alpha} \in \{0,1\}^p$, we may rewrite the trust-region sub-problem (Hamming distance) constraint as:
\begin{equation*}
     \|\boldsymbol{\alpha}-\boldsymbol{\alpha}^{(k)}\|_1 =\sum_{\substack{j=1 \\\alpha_j^{(k)}=0}}^p\alpha_j +  \sum_{\substack{j=1 \\\alpha_j^{(k)}=1}}^p( 1-\alpha_j) \leq \Delta_k,
\end{equation*}
which is linear in $\alphab$. Hence, the trust-region sub-problem is a knapsack problem with computational complexity $\mathcal{O}(p\log(p))$ \cite{dudzinski1987exact}. In the context of this work, the computational complexity for evaluating the gradient is $\mathcal{O}(p)$. This means that the first-order trust-region method applied to the SFQ problem has computational complexity $\mathcal{O}(p\log(p))$. 

The convergence of the binary trust-region is inherited from the fact that the standard continuous trust-region method converges monotonically, implying that the objective function value must remain the same or decrease after each trust-region iteration \cite{nocedal.wright:99}. This is to say that at step $k$ of the standard trust-region method:
\begin{align*}
    \mathcal{J}(\boldsymbol{\alpha}_{k+1}) \leq \mathcal{J}(\boldsymbol{\alpha}_k). 
\end{align*}
In our context, a binary (local) optimal solution $\tilde{\boldsymbol{\alpha}}$ satisfies the condition:
\begin{align*}
     \mathcal{J}(\tilde{\boldsymbol{\alpha}}) \leq \mathcal{J}(\boldsymbol{\alpha}), 
\end{align*}
for all $\boldsymbol{\alpha}$ satisfying
\begin{align*}
    ||\tilde{\boldsymbol{\alpha}} - \boldsymbol{\alpha}||_1\leq 1.
\end{align*}
This is to say no local neighbor in binary space under the Hamming measure produces a lower objective function value. In this work we assume the existence of at least one local minimizing solution $\tilde{\boldsymbol{\alpha}}$ (even though it may not be the only local optimal solutions), which the trust-region is theoretically guaranteed to converge to \cite{nocedal.wright:99}, and indeed find such minimum numerically. A proof for the existence of such minimizing solutions is out of the scope of this paper.

As we proceed to show in Section \ref{s:experiements}, our algorithm finds quality local solutions even if the binary optimization space is of high dimension. We note that we chose a first-order trust-region method versus a second order trust-region method because then the trust-region problem would become a quadratic integer programming problem, which could be computationally intractable.

\section{Numerical experiments}\label{s:experiements}
To evaluate the performance of our trust region algorithm, we consider optimizing pulse sequences for two standard quantum gates: H and X. For each of these gates we study the cases when the pulse strength $\beta$ corresponds to the tip angles $\theta=\frac{\pi}{100}$ and $\theta=\frac{\pi}{300}$. 

To model a typical superconducting transmon, we set the fundamental frequency, self-Kerr coefficient, SFQ time step, and pulse duration according to
\begin{gather*}
\frac{\omega}{2\pi} =  5.0 \ \mbox{GHz},\quad
\frac{\xi}{2\pi} = 0.25 \ \mbox{GHz}\quad
\tau_p = 2.5\times 10^{-2} \ \mbox{ns}, \\ 
\delta=4 \times 10^{-3}  \ \mbox{ns}.
\end{gather*}
We numerically calculate the unitary matrices $D_0$ and $D_1$ by the Str\"omer-Verlet method, which is a partitioned Runge-Kutta scheme \cite{anders2020discrete}. To effectively eliminate time-stepping errors, we solve the state equation on the interval $[0,\tau_p]$ with 10,000 time steps. Truncating the modal expansion of Schr\"odinger's equation after $N$ terms is only valid if the occupation of the highest energy level is sufficiently small. From numerical experimentation we find that retaining four energy levels is sufficient for meeting that requirement; in the following, we use 2 essential states ($E=2$) and 2 guard states ($G=2$). The essential states correspond to the number states $\myket{0}$ and $\myket{1}$, and the guard states are $\myket{2}$ and $\myket{3}$. The truncation of the modal expansion is justified when the occupation of the $\myket{3}$ state is small throughout the duration of the pulse sequence. Furthermore, for all experiments, we take $w_1=0.1$ and $w_2=1$ in the weight matrix $W$. In our experiments we observe that larger values of $w_1$ and $w_2$ reduce the leakage to higher energy states, but leads to a larger infidelity. We note that the values for $w_1$ and $w_2$ we use here are only tuned by numerical experiments. They may be further improved through a separate optimization procedure. We start by fixing the duration of each gate to $T=40$ ns, which correspond to $p=1,600$ SFQ time steps. In each experiment we apply the trust-region method for 10 random binary initial guesses, and report the trust-region solution that resulted in the smallest objective function. In these experiments we set the acceptance step size to be $\hat{\rho}=0.75$, and the initial trust-region radius to equal the number of SFQ pulses ($\Delta_0=p)$. In a separate set of numerical experiments, reported in Section~\ref{s:minimum}, we explore the shortest gate duration that is needed to realize the gates to high fidelity.

\subsection{H gate experiment}
For an H gate, the target unitary matrix is
\begin{equation*} \label{hgate}
V_E = \frac{1}{\sqrt{2}}\begin{bmatrix}
1 & 1 \\
1 & -1
\end{bmatrix}.
\end{equation*}
Starting from the $\myket{0}$ or $\myket{1}$ states, an H gate is realized when the probabilities of being in either the $\myket{0}$ or $\myket{1}$ states become equal. 
\begin{figure}[htbp]
\centering
\subfloat[$\theta = \frac{\pi}{300}$.]{\includegraphics[width=0.45\textwidth]{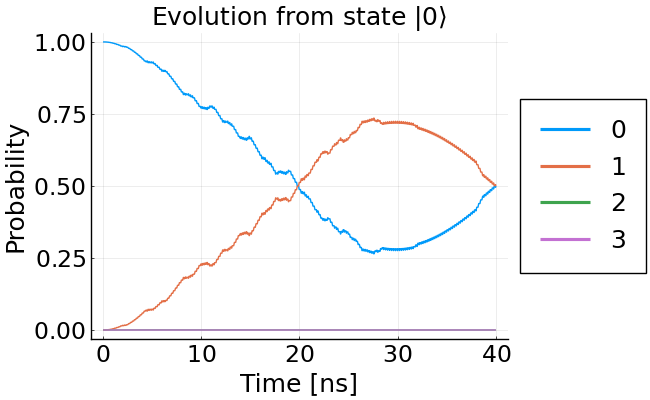}}\hspace{5mm}
\subfloat[$\theta = \frac{\pi}{100}$.]{\includegraphics[width=0.45\textwidth]{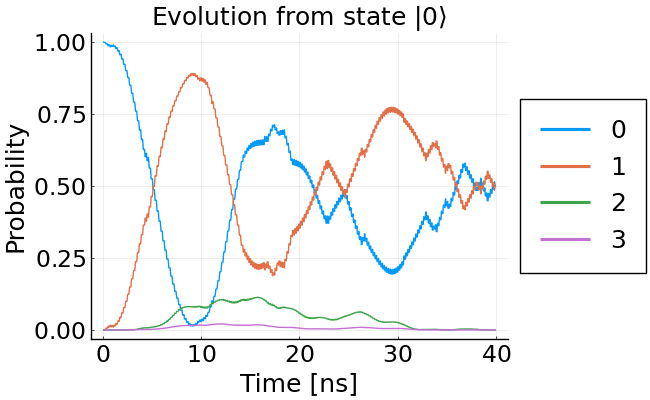}}\\
\subfloat[$\theta = \frac{\pi}{300}$.]{\includegraphics[width=0.45\textwidth]{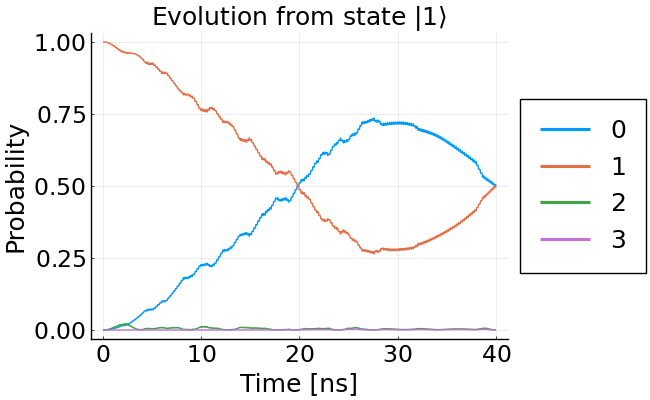}}\hspace{5mm}
\subfloat[$\theta = \frac{\pi}{100}$.]{\includegraphics[width=0.45\textwidth]{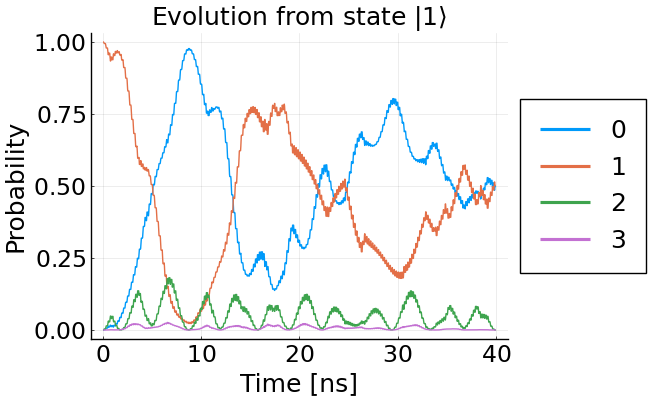}}\\
\subfloat[$\theta = \frac{\pi}{300}$.]{\includegraphics[width=0.45\textwidth]{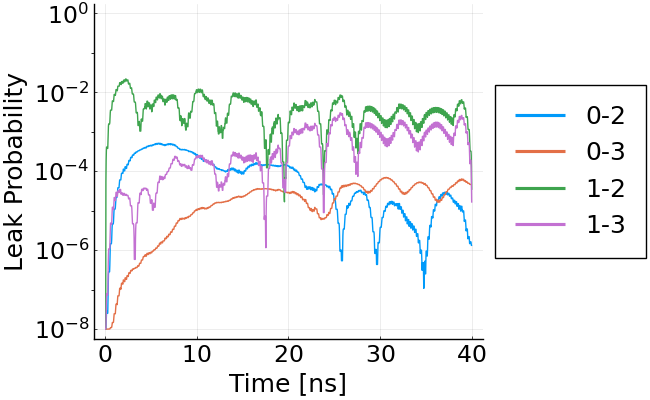}}\hspace{5mm}
\subfloat[$\theta = \frac{\pi}{100}$.]{\includegraphics[width=0.45\textwidth]{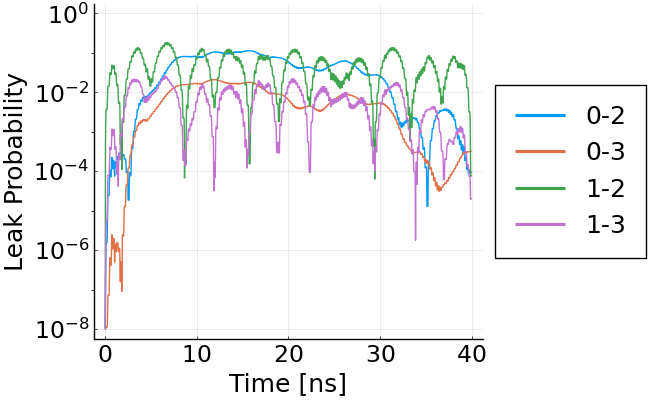}} \\
\subfloat[Pulse sequence for $\theta = \frac{\pi}{300}$.]{\includegraphics[width=0.45\textwidth]{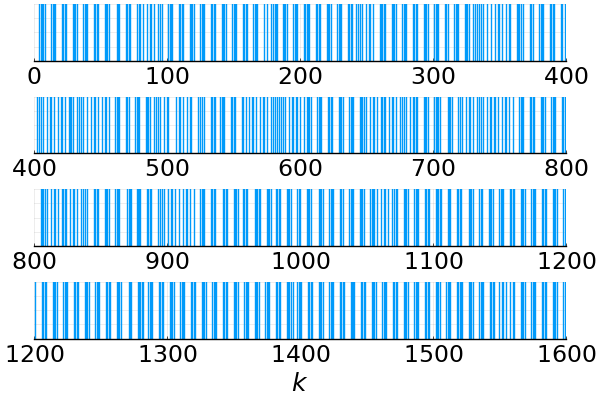}}\hspace{5mm}
\subfloat[Pulse sequence for $\theta = \frac{\pi}{100}$.]{\includegraphics[width=0.45\textwidth]{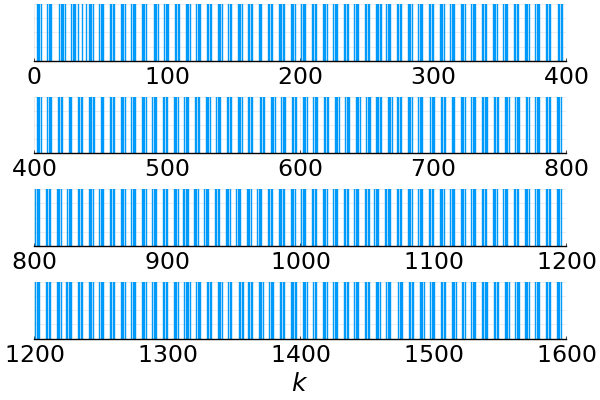}}

\caption{The evolution of essential and guard state populations during the H gate with tip angle $\theta = \frac{\pi}{300}, \frac{\pi}{100}$, and corresponding pulse sequences. In figures (e) and (f), the legend $a-b$ indicates the population of $\myket{b}$ corresponding to initial state $\myket{a}$. }
\label{fig:hgate}
\end{figure}

In Figure \ref{fig:hgate} we plot the populations for the essential states, the leakage from the essential to the guard states, and the corresponding pulse sequence, for tip angles $\frac{\pi}{300}$ and $\frac{\pi}{100}$. We observe that we are successful in realizing the H gate for both tip angles when the duration is $T=40$ ns. In the case of tip angle $\theta=\frac{\pi}{300}$ we observe an infidelity that is less than $10^{-4}$, corresponding to a gate fidelity larger then $99.99$ \%. For the tip angle $\theta=\frac{\pi}{100}$ we observe an infidelity less than $10^{-3}$. In that case, the gate fidelity exceeds $99.9\%$. Thus both tip angles meet the gate fidelity threshold of 99.9 \%.

A desired trait of our approach is that the leakage from the essential states to the guard states is suppressed. The area under the delivered pulse, $\beta$, can be physically interpreted as the energy invested into the system. Because $\beta \sim \theta$, a scaling in $\theta$ corresponds to an equivalent scaling in $\beta$. So the energy delivered by a pulse with tip angle $\theta=\frac{\pi}{100}$ is three times stronger than the energy delivered by a pulse with tip angle $\theta =\frac{\pi}{300}$. We observe in the leakage plots in Figure~\ref{fig:hgate} that more leakage occurs for the larger tip angle, because that corresponds to administering a pulse with more energy. This indicates that there is practical impact of using higher energy pulses compared to lower energy pulses. This also motivates the introduction of the leakage term $\mathcal{J}_2$ to suppress transition to the guard states. To justify the truncation of the modal expansion, it is desired to see sufficiently small population of the $\myket{3}$ state. In this case the largest population of $\myket{3}$ is less than $10^{-2}$ for $\theta =\frac{\pi}{300}$, and less than $10^{-1}$ for $\theta=\frac{\pi}{100}$. This observation indicates that there is advantage to using weaker pulses, because it leads to low infidelity and smaller population of the $\myket{3}$ state. Furthermore, the stronger pulse strength cause more leakage to occur, which increases the leakage term ($\mathcal{J}_2$) relative to the infidelity term (${\cal J}_1$) in the objective function. As a result, the infidelity becomes an order of magnitude larger compared to a weaker pulse. This suggests that stronger pulses may not be favored. In Section \ref{s:minimum} we discuss how stronger pulses, while generating more leakage, allow the gate duration to be reduced without increasing the gate infidelity.

In Figure \ref{fig:hgate} we present the pulse sequences for each of the two tip angles. We plot these in a barcode format, where the presence of a line indicates $\alpha_k=1$ for a given $k$ and otherwise $\alpha_k=0$. We observe that there is a structure that emerges from solving the optimal control problem, with packets of pulses being created in what seems to be a semi-periodic structure. 

\begin{figure}[htbp]
\centering
\subfloat[$\theta = \frac{\pi}{300}$.]{\includegraphics[width=0.45\textwidth]{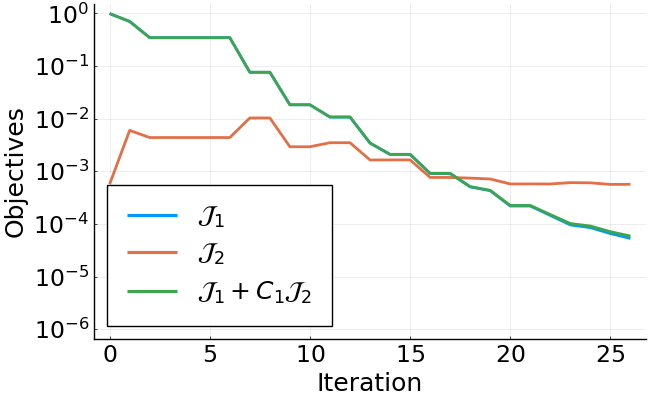}}\hspace{3mm}
\subfloat[$\theta = \frac{\pi}{100}$.]{\includegraphics[width=0.45\textwidth]{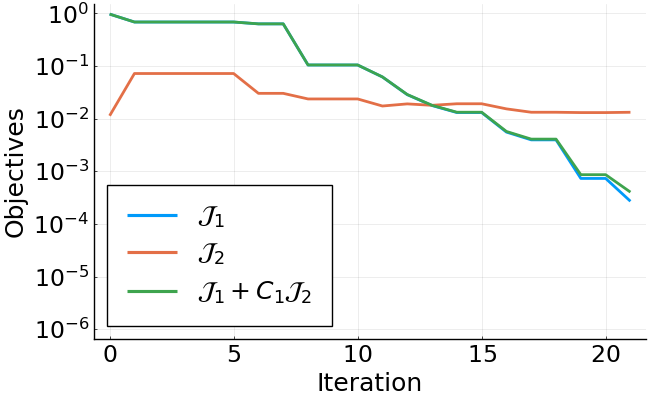}}

\caption{Convergence history for the trust-region method for realizing an H gate with 1600 control pulses, for tip angles $\theta = \frac{\pi}{300}$ (left) and $\frac{\pi}{100}$ (right). }
\label{fig:trh}
\end{figure}

In Figure \ref{fig:trh} we plot, for $\theta= \frac{\pi}{300}$ and $\frac{\pi}{100}$, the trust-region convergence history of the objective functions $\mathcal{J}_1$, $\mathcal{J}_2$, and $\mathcal{J}_1 + C_1\mathcal{J}_2$. These plots illustrate that the number of iterations required to find a local minimizing binary solution is modest. For $\theta = \frac{\pi}{300}$, we find a minimal solution in around 25 iterations. In the case of $\theta = \frac{\pi}{100}$, around 20 iterations are needed. Since the number of control pulses in this case are $p=1600$, there are $2^{1600} \approx 4.44\times 10^{481}$ feasible solutions to the binary optimal control problem. In spite of the enormous size of the solution space, our method quickly finds quality solutions for both tip angles, with significant reduction in the infidelity term. We also observe that the leakage is small in the initial guess and does not become larger in the optimized solution. Similar convergence histories are observed for the X gate as well.

\subsection{X gate experiment}
An X gate corresponds to the unitary transformation 
\begin{equation*} \label{xgate}
V_E = \begin{bmatrix}
0 & 1 \\
1 & 0
\end{bmatrix}.
\end{equation*}
To realize the X gate is equivalent to saying that the populations of the initial states are swapped, e.g. if the qubit starts in state $\myket{0}$, it will transition to state $\myket{1}$, and vice versa.
\begin{figure}[htbp]
\centering
\centering
\subfloat[$\theta = \frac{\pi}{300}$.]{\includegraphics[width=0.45\textwidth]{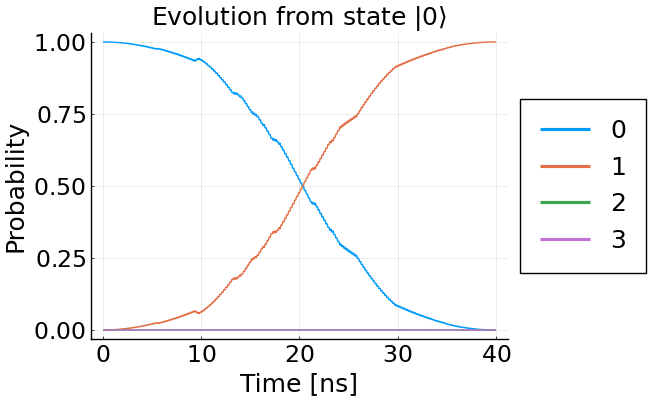}}\hspace{5mm}
\subfloat[$\theta = \frac{\pi}{100}$.]{\includegraphics[width=0.45\textwidth]{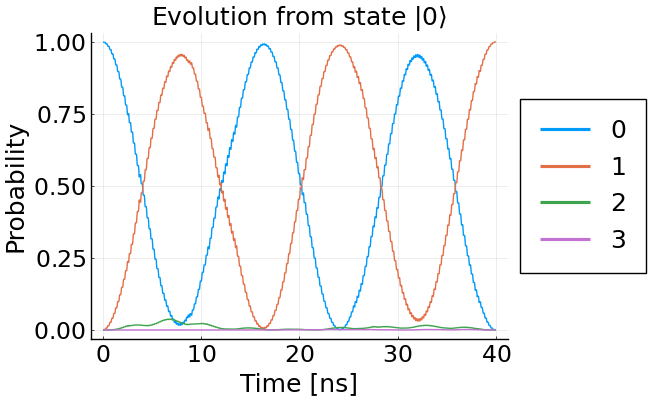}}\\
\subfloat[$\theta = \frac{\pi}{300}$.]{\includegraphics[width=0.45\textwidth]{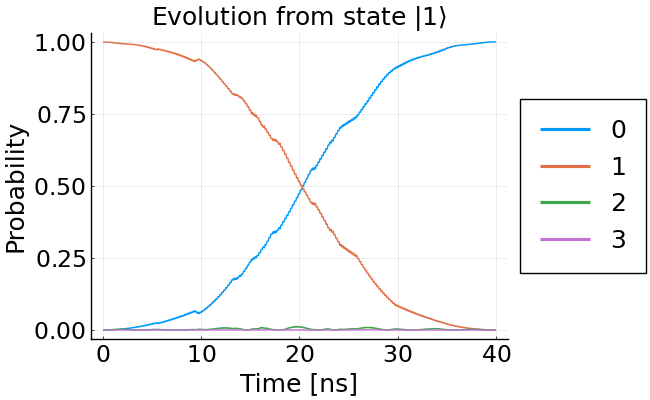}}\hspace{5mm}
\subfloat[$\theta = \frac{\pi}{100}$.]{\includegraphics[width=0.45\textwidth]{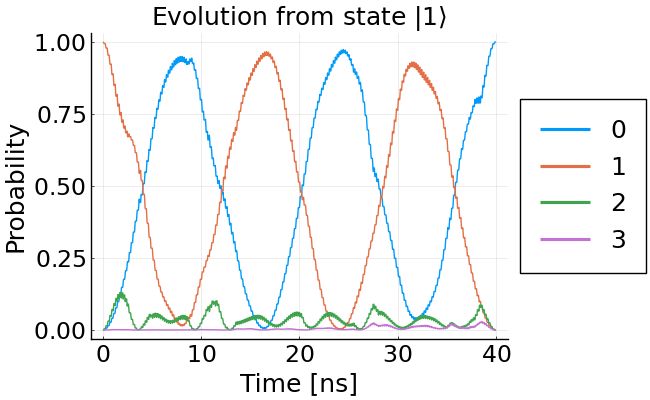}}\\
\subfloat[$\theta = \frac{\pi}{300}$.]{\includegraphics[width=0.45\textwidth]{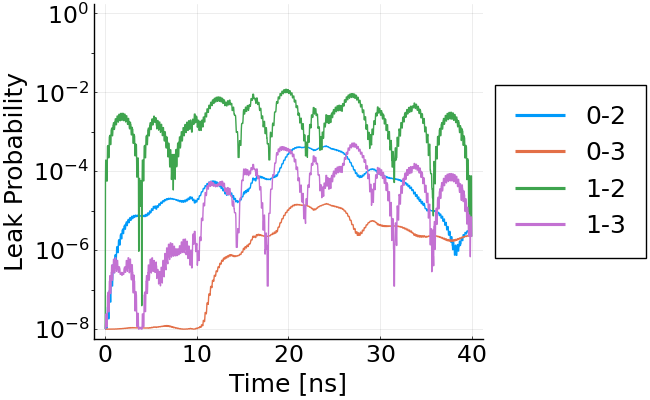}}\hspace{5mm}
\subfloat[$\theta = \frac{\pi}{100}$.]{\includegraphics[width=0.45\textwidth]{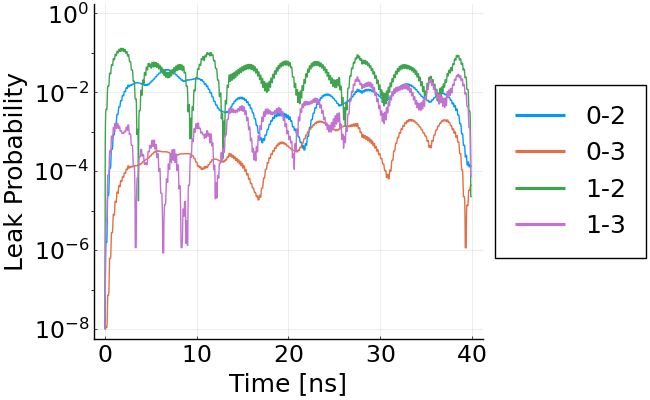}} \\
\subfloat[Pulse sequence for $\theta = \frac{\pi}{300}$.]{\includegraphics[width=0.45\textwidth]{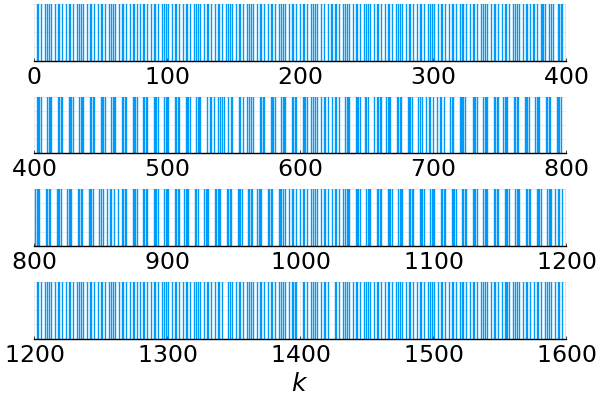}}\hspace{5mm}
\subfloat[Pulse sequence for $\theta = \frac{\pi}{100}$.]{\includegraphics[width=0.45\textwidth]{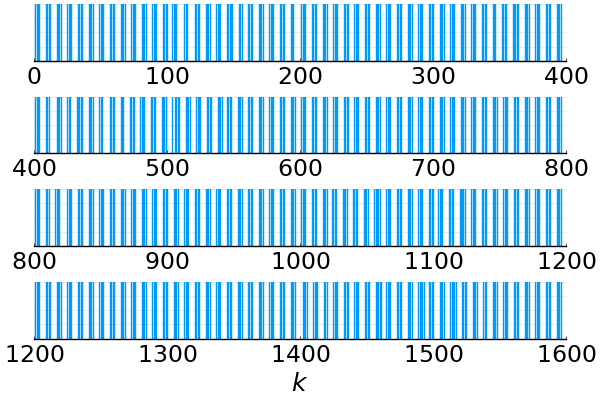}}

\caption{The evolution of essential and guard state populations during the X gate, with tip angles $\theta = \frac{\pi}{300}$ and $\frac{\pi}{100}$, and corresponding pulse sequences. In figures (e) and (f), the legend $a-b$ indicates the population of $\myket{b}$ corresponding to initial state $\myket{a}$.}
\label{fig:xgate}
\end{figure}

In Figure \ref{fig:xgate} we show the populations for the essential states, the leak probability from the essential to the guard states, and the pulse sequences associated with the tip angles $\theta = \frac{\pi}{300}$ and $\frac{\pi}{100}$. We observe close similarities between the H and X gates. First, we are able to realize the gates with infidelity that is less than $10^{-4}$ for tip angle $\theta = \frac{\pi}{300}$. In addition we observe an infidelity less than $10^{-3}$ for a tip angle of $\theta = \frac{\pi}{100}$. We observe more leakage for a stronger pulse than a weaker pulse. Similar to the H gate, there is a semi-periodic structure in the X gate pulse sequences.

\subsection{Comparison with the branch and bound algorithm}
The branch and bound (B$\&$B) method is the primary algorithm leveraged for integer programming problems. To read more about the standard method we refer the reader to \cite{bonami2011more}. B$\&$B only has theoretical guarantees of finding an optimal solution, even though it may take enormous effort, in the case of convex integer programming problems. In the case of nonconvex problems, like our SFQ optimal control problem, the method is merely a heuristic. To compare with our trust-region method, we applied the B$\&$B approach to the H and X gate experiments. First, we executed the B$\&$B algorithm using several different branching strategies: most infeasible branching, pseudo cost branching, and strong branching. For each branching strategy we allowed the  B$\&$B solution to search for a solution for the duration of one hour. In each run,  B$\&$B not only could not find an integer optimal solution, but the quality of the fidelity for both the H and X gate was only on the order of $10^{-1}$. We conducted the experiment again, this time increasing the search time allotted from one hour to ten hours. The quality of the infidelity remained unchanged. These experiments indicate that the standard branch and bound approach is not successful at finding a quality solution; whereas our trust-region approaches finds solutions with infidelity less than $10^{-3}$ within one minute of computation. This isn't to say however that a specialized  B$\&$B scheme could not be developed for the SFQ application, however to develop such a scheme would require deep expertise in the theory of integer programming, and is still not guaranteed to solve the SFQ optimal control problem. While an interesting research direction, developing a specialized  B$\&$B for this application is out of the scope of the paper. We refer the reader to convexification \cite{c3,c1,c2} and underestimators  \cite{u1,u3,u2} used in order to develop specialized B$\&$B schemes for nonconvex integer programming, to give insight how one may begin to think about developing  a specialized  B$\&$B scheme for this application.

\subsection{On minimizing the gate duration}\label{s:minimum}

We now examine the smallest gate duration that is required to achieve an acceptable gate infidelity for the H and X gates. 

\begin{figure}[htbp]
\centering
\subfloat[ H gate $\theta = \frac{\pi}{300}$.]{\includegraphics[width=0.45\textwidth]{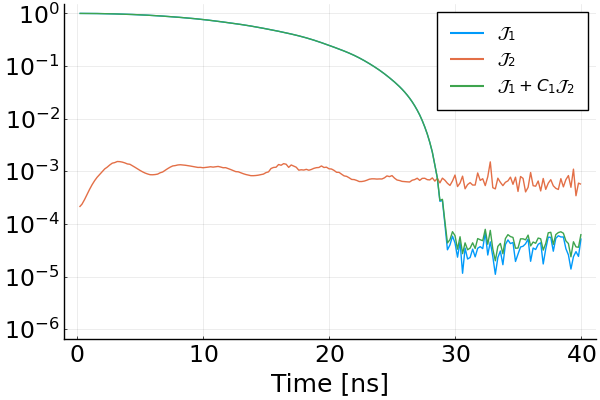}}\hspace{5mm}
\subfloat[H gate $\theta = \frac{\pi}{100}$.]{\includegraphics[width=0.45\textwidth]{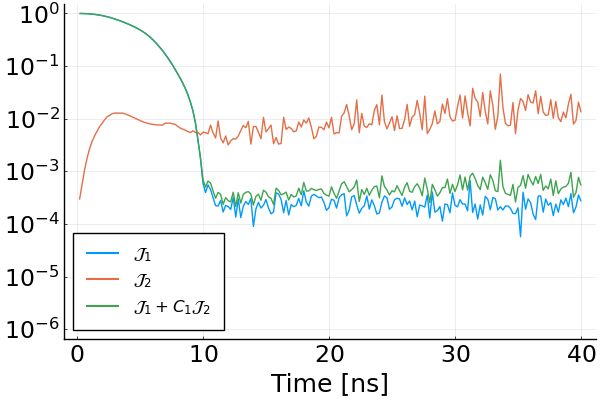}} \\
\subfloat[X gate $\theta = \frac{\pi}{300}$,]{\includegraphics[width=0.45\textwidth]{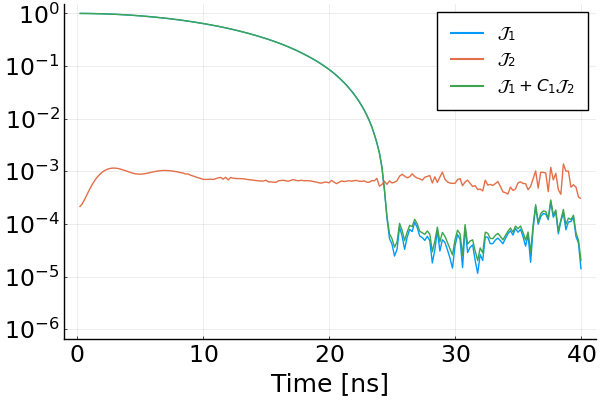}}\hspace{5mm}
\subfloat[X gate $\theta = \frac{\pi}{100}$]{\includegraphics[width=0.45\textwidth]{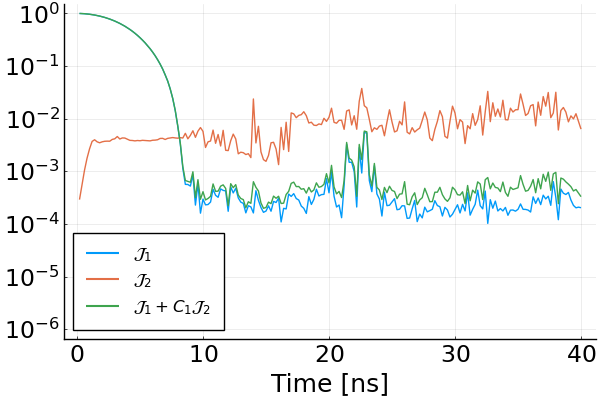}}
\caption{The infidelity, leakage, and total objective function on a $\log_{10}$ scale for tip angle $\theta = \frac{\pi}{300},\frac{\pi}{100}$ for the H and X gates.}
\label{fig:fidofgate}
\end{figure}

In Figure \ref{fig:fidofgate} we study the behavior of the infidelity as a function of gate duration. We solve the optimal control problem for a gate duration of $[0,\,p\tau_p]$ for $p=8,\,16,\,24,\,\ldots,1600$. For each value of $p$, we draw 10 random samples from $\{0,1 \}^p$ as initial guesses, and execute the trust-region on each sample. The trust-region solution with the lowest objective function value is shown in Figure \ref{fig:fidofgate}.
For all of the gates and tip angles considered here, we find solutions with an infidelity less than $10^{-3}$ for gate durations less than 40 ns. We observe that a pulse with three times the strength produces an infidelity that is less than $10^{-3}$ for a gate duration that is about three times shorter. In particular, for the H gate we realize a gate infidelity that is less than $10^{-3}$ for the tip angle $\theta = \frac{\pi}{300}$ in $\approx 28$ ns, versus $\approx 10$ ns for the tip angle $\theta = \frac{\pi}{100}$. For the X gate, we realize a gate infidelity that is less than $10^{-3}$ for the tip angle $\theta = \frac{\pi}{300}$ in $\approx 24$ ns, versus $\approx 8$ ns for tip angle $\theta = \frac{\pi}{100}$. 

 Our experiments indicate that high fidelity gates can be realized in shorter duration than 40 ns. The leakage to the $\myket{3}$ state largely depends on the amplitude of each pulse, i.e., less leakage is achieved by decreasing the tip angle. To find a solution that meets the needs for a small infidelity and a small leakage, we suggest an iterative approach where the number of pulses is gradually increased until the infidelity becomes acceptable. If the leakage is too large, reduce the tip angle and repeat the calculation for a larger number of pulses.

\section{Conclusions}\label{s:conclusions}
In this paper we introduce a novel approach for optimizing pulse delivery to a SFQ quantum computer. We pose the problem as a binary optimal control problem, where the binary variables indicate whether each pulse is on or off, with a primary goal of minimizing the gate infidelity. A secondary goal of our formulation is to suppress leakage to the higher energy levels. To solve the optimal control problem we take advantage of a first order trust-region method. We devise an algorithm for calculating the objective function and its relaxed gradient in ${\cal O}(p)$ operations, where $p$ is the number of SFQ time steps. Based on the linearity of the Hamming distance that appears in the trust-region sub-problem, the overall computational complexity of the algorithm becomes $\mathcal{O}(p \log p)$. Our numerical experiments indicate that our trust-region method needs $\approx 25$ iterations to find local minima with gate fidelities better than an $99.9\%$. We also investigated the minimal gate duration that is required to realize a gate. The actual duration varies from gate to gate, but we have demonstrated that we are able to find pulse sequences for the H and X gate. Furthermore, our numerical experiments indicate pros and cons when it comes to using stronger versus weaker pulses. For a given gate, we observe that increasing the pulse amplitude by a factor of three allows the gate duration to be reduced by a factor of three. However, using stronger pulses results in more leakage, which is not desired. When using weaker pulses, we observe less leakage, but the gate duration must be increased to achieve an acceptable gate fidelity. We note that all behaviors we have observed for the H and X gates have also been observed in separate numerical experiments for the Y and Z gates. Those experiments were omitted to conserve space.

This paper illustrates that the reduced gradient trust region method is a very promising candidate for optimizing SFQ pulse sequences. Initial numerical experiments indicate that our approach will generalize in a straightforward way to SFQ control of two-qubit systems. While out of the scope of this work, in a future work we intend to introduce the final time duration $T$ as an optimization variable; and penalize it such that we realize gates in the shortest duration possible for both single and many qubit systems.

\section*{Acknowledgments}
This document was prepared as an account of work sponsored by an agency of the
United States government. Neither the United States government nor Lawrence
Livermore National Security, LLC, nor any of their employees makes any warranty,
expressed or implied, or assumes any legal liability or responsibility for the accuracy,
completeness, or usefulness of any information, apparatus, product, or process disclosed, or represents that its use would not infringe privately owned rights. Reference herein to any specific commercial product, process, or service by trade name, trademark,
manufacturer, or otherwise does not necessarily constitute or imply its endorsement,
recommendation, or favoring by the United States government or Lawrence Livermore
National Security, LLC. The views and opinions of authors expressed herein do not
necessarily state or reflect those of the United States government or Lawrence Livermore
National Security, LLC, and shall not be used for advertising or product endorsement
purposes. This work was performed under the auspices of the U.S. Department of Energy by Lawrence Livermore National Laboratory under Contract DE-AC52-07NA27344; this is contribution LLNL-JRNL-823366.

\appendix
\section{Gradient computation} \label{s:gradcomp}
The gradient of $U_j$ with respect to $\alpha_k$ satisfies
\begin{align}\label{eq_3}
    \frac{\partial U_j}{\partial \alpha_k} =  
    \begin{cases}
        0, & k>j,\\
        B_k U_{k-1}, & k=j,\\
        A_{k+1} B_k U_{k-1}, & k = j-1,\\
        A_j \cdots A_{k+1} B_k U_{k-1}, &  k\leq j-2,\\
        A_j \cdots A_{2} B_1, &  k = 1.
    \end{cases}
\end{align}
Recall that $p$ corresponds to the total number of control pulses and that $S_T\in\mathbb{C}$. By differentiating \eqref{eq_J1} with respect to $\alpha_p$ and using \eqref{eq_3}, we get
\begin{align*}
    \frac{d {\cal J}_1}{d \alpha_p} = -\frac{2}{E^2} \mbox{Re} \left( \bar{S}_T \left\langle \frac{d U_p}{d \alpha_p} P, V P\right\rangle_F 
    \right) 
    = 
    -\frac{2}{E^2} \mbox{Re} \left( \bar{S}_T \left\langle B_p U_{p-1} P, V P\right\rangle_F \right).
\end{align*}
For $k=p-1$, we observe
\begin{align*}
    \frac{d {\cal J}_1}{d \alpha_{p-1}} &= -\frac{2}{E^2} \mbox{Re} \left( \bar{S}_T \left\langle \frac{d U_p}{d \alpha_{p-1}} P, V P\right\rangle_F 
    \right) \\
    &= 
    -\frac{2}{E^2} \mbox{Re} \left( \bar{S}_T \left\langle A_p B_{p-1} U_{p-2} P, V P\right\rangle_F \right)\\
    &= 
    -\frac{2}{E^2} \mbox{Re} \left( \bar{S}_T \left\langle  B_{p-1} U_{p-2} P, A^\dagger_p V P\right\rangle_F \right).
\end{align*}
For $k=p-2$, we observe
\begin{align*}
    \frac{d {\cal J}_1}{d \alpha_{p-2}} &= -\frac{2}{E^2} \mbox{Re} \left( \bar{S}_T \left\langle \frac{d U_p}{d \alpha_{p-2}} P, V P\right\rangle_F 
    \right) \\
    &= 
    -\frac{2}{E^2} \mbox{Re} \left( \bar{S}_T \left\langle A_p A_{p-1} B_{p-2} U_{p-3} P, V P\right\rangle_F \right)\\
    &= 
    -\frac{2}{E^2} \mbox{Re} \left( \bar{S}_T \left\langle  B_{p-2} U_{p-3} P, A^\dagger_{p-1} A^\dagger_p  V P\right\rangle_F \right).
\end{align*}
Let's define the discrete adjoint variable $\Lambda_k$ according to
\begin{align*}
    \Lambda_p &= V,\\
    \Lambda_{p-1} & = A^\dagger_p V
    = A_p^\dagger \Lambda_p,\\
    \Lambda_{p-2} & = A^\dagger_{p-1} A^\dagger_p  V
    = A^\dagger_{p-1} \Lambda_{p-1},\\
    &\vdots\nonumber \\
    \Lambda_{p-q} & = A^\dagger_{p-q+1} \Lambda_{p-q+1}
\end{align*}
This means that all components of the gradient can be evaluated by backwards propagation of the adjoint variable,
\begin{align*}
    \frac{d {\cal J}_1}{d \alpha_{p-q}} = -\frac{2}{E^2} \mbox{Re} \left( \bar{S}_T \left\langle B_{p-q} U_{p-q-1} P, \Lambda_{p-q} P\right\rangle_F 
    \right), \quad q=0, 1, \ldots, p-1.
\end{align*}

Note that $U_0 = I$ does not depend on $\boldsymbol{\alpha}$. Because $W^\dagger = W$, the gradient of the leak term satisfies
\begin{align*}
    \frac{d{\cal J}_2}{d\alpha_k} &= 
        \frac{2}{p} \mbox{Re} \left(
     \sum_{j=1}^{p-1}\left\langle 
     \frac{d U_j}{d \alpha_k}P, W U_jP 
     \right\rangle_F
    + \frac{1}{2}  
    \left\langle 
     \frac{d U_p}{d \alpha_k}P, W U_pP 
     \right\rangle_F
    \right) \nonumber \\
    &=
     \frac{2}{p} \mbox{Re} \left(
     \sum_{j=k}^{p-1}\left\langle 
     \frac{d U_j}{d \alpha_k}P, W U_jP 
     \right\rangle_F
    + \frac{1}{2}  
    \left\langle 
     \frac{d U_p}{d \alpha_k}P, W U_pP 
     \right\rangle_F
    \right)
\end{align*}
For $k=p$,
\begin{align*}
    \frac{d{\cal J}_2}{d\alpha_p} &= 
        \frac{2}{p} \mbox{Re} 
    \left\langle 
     \frac{d U_p}{d \alpha_p}P,\frac{1}{2} W U_pP 
     \right\rangle_F =
     \frac{2}{p} \mbox{Re} 
    \left\langle 
     B_p U_{p-1}P,\frac{1}{2} W U_p P
     \right\rangle_F
\end{align*}
For $k=p-1$,
\begin{align*}
    \frac{d{\cal J}_2}{d\alpha_{p-1}}&=
    \frac{2}{p} \mbox{Re} \left(
     \left\langle 
     \frac{d U_{p-1}}{d \alpha_{p-1}}P, W U_{p-1}P 
     \right\rangle_F
    +   
    \left\langle 
     \frac{d U_p}{d \alpha_{p-1}}P, \frac{1}{2} W U_pP 
     \right\rangle_F
    \right) \nonumber \\
    &=
   \frac{2}{p} \mbox{Re} \left(
     \left\langle 
     B_{p-1}U_{p-2}P, W U_{p-1}P 
     \right\rangle_F
    +   
    \left\langle 
     A_p B_{p-1} U_{p-2}P, \frac{1}{2} W U_p P 
     \right\rangle_F
    \right) \nonumber \\
    &=
    \frac{2}{p} \mbox{Re} \left(
     \left\langle 
     B_{p-1}U_{p-2} P, W U_{p-1}P + \frac{1}{2} A^\dagger_p W U_p P
     \right\rangle_F
    \right)
\end{align*}
For $k=p-2$,
\begin{align*}
    \frac{d{\cal J}_2}{d\alpha_{p-2}}&=
    \frac{2}{p} \mbox{Re} \bigg(
     \left\langle 
     \frac{d U_{p-2}}{d \alpha_{p-2}} P, W U_{p-2} P 
     \right\rangle_F +
     \left\langle 
     \frac{d U_{p-1}}{d \alpha_{p-2}} P, W U_{p-1} P 
     \right\rangle_F\\
    &+   
    \left\langle 
     \frac{d U_p}{d \alpha_{p-2}} P, \frac{1}{2} W U_p P 
     \right\rangle_F
    \bigg)\nonumber \\
    &=
    \frac{2}{p} \mbox{Re} \bigg(
     \left\langle 
     B_{p-2}U_{p-3} P, W U_{p-2} P 
     \right\rangle_F +
     \left\langle 
     A_{p-1}B_{p-2} U_{D-3} P, W U_{D-1}P 
     \right\rangle_F\\
    &+   
    \left\langle 
    A_D A_{p-1}B_{p-2}U_{p-3}P, \frac{1}{2} W U_p P 
     \right\rangle_F
    \bigg)\nonumber \\
    &=
    \frac{2}{p} \mbox{Re} 
     \left\langle 
     B_{p-2}U_{p-3}, W U_{p-2} + A^\dagger_{p-1}W U_{p-1} + \frac{1}{2} A^\dagger_{p-1} A^\dagger_p W U_p 
     \right\rangle_F 
\end{align*}
We now define
\begin{align*}
    \Tilde{\Lambda}_p &= \frac{1}{2} W U_p P,\\
    \Tilde{\Lambda}_{p-1} & = W U_{p-1}P + \frac{1}{2} A^\dagger_p W U_pP = W U_{p-1}P + A_p^\dagger \widetilde{\Lambda}_p,\\
    \Tilde{\Lambda}_{p-2} & = W U_{p-2}P + A^\dagger_{p-1}W U_{p-1}P + \frac{1}{2} A^\dagger_{p-1} A^\dagger_p W U_pP = W U_{p-2}P + A^\dagger_{p-1} \widetilde{\Lambda}_{p-1}
\end{align*}
One can verify the recursive formula,
\begin{align*}
   \Tilde{\Lambda}_{p-q} = \begin{cases}
    \frac{1}{2} W U_p P,& q=0,\\
        W U_{p-q}P + A^\dagger_{p-q+1} \widetilde{\Lambda}_{p-q+1},&q = 1,2,\ldots,p-1,
    \end{cases}
\end{align*}
from which all components of the gradient of the objective function can be calculated.
As a result, the contribution to the gradient from the leak term becomes
\begin{align*}
    \frac{d{\cal J}_2}{d\alpha_{p-q}} = 
    \frac{2}{p} \mbox{Re} 
     \left\langle 
     B_{p-q}U_{p-q-1}P,\Tilde{\Lambda}_{p-q}P \right\rangle_F,\quad q=0,1,\ldots, p-1.
\end{align*}

\bibliography{references}

\end{document}